# Electronic structure and direct observation of ferrimagnetism in multiferroic hexagonal YbFeO$_3$

Shi Cao,[1] Kishan Sinha,[1] Xin Zhang,[1] Xiaozhe Zhang,[1,2] Xiao Wang,[3] Yuewei Yin,[1] Alpha T N'Diaye,[4] Jian Wang,[5] David J Keavney,[6] Tula R Paudel,[1] Yaohua Liu,[7] Xuemei Cheng,[3] Evgeny Y Tsymbal,[1,8] Peter A Dowben,[1,8] Xiaoshan Xu[1,8]

[1]Department of Physics and Astronomy, University of Nebraska, Lincoln, NE 68588, USA

[2]Department of Physics, Xi'an Jiaotong University, Xi'an 710049, People's Republic of China

[3]Department of Physics, Bryn Mawr College, Bryn Mawr, PA 19010, USA

[4]Advanced Light Source, Lawrence Berkeley National Laboratory, Berkeley, CA 94720, USA

[5]Canadian Light Source, Saskatoon, SK S7N 2V3, Canada

[6]Advanced Photon Source, Argonne National Laboratory, Argonne, Illinois 60439, USA

[7]Quantum Condensed Matter Division, Oak Ridge National Lab, Oak Ridge, TN 37831, USA

[8]Nebraska Center for Materials and Nanoscience, University of Nebraska, Lincoln, NE 68588, USA

## Abstract

The magnetic interaction between rare-earth and Fe ions in hexagonal rare-earth ferrites (h-REFeO$_3$), may amplify the weak ferromagnetic moment on Fe, making these materials more appealing as multiferroics. To elucidate the interaction strength between the rare-earth and Fe ions as well as the magnetic moment of the rare-earth ions, element specific magnetic characterization is needed. Using X-ray magnetic circular dichroism, we have studied the ferrimagnetism in h-YbFeO$_3$ by measuring the magnetization of Fe and Yb separately. The results directly show anti-alignment of magnetization of Yb and Fe ions in h-YbFeO$_3$ at low temperature, with an exchange field on Yb of about 17 kOe. The magnetic moment of Yb is about 1.6 $\mu_B$ at low-temperature, significantly reduced compared with the 4.5 $\mu_B$ moment of a free Yb$^{3+}$. In addition, the saturation magnetization of Fe in h-YbFeO$_3$ has a sizable enhancement compared with that in h-LuFeO$_3$. These findings directly demonstrate that ferrimagnetic order exists in h-YbFeO$_3$; they also account for the enhancement of magnetization and the reduction of coercivity in h-YbFeO$_3$ compared with those in h-LuFeO$_3$ at low temperature, suggesting an important role for the rare-earth ions in tuning the multiferroic properties of h-REFeO$_3$.

# I. Introduction

The diverse magnetic properties of rare-earth (RE) transition-metal (TM) oxides owe to the interplay between the distinct magnetism of rare-earth and transition-metal ions. For the transition-metal ions, the magnetic moments come from d electrons which are well exposed to the local environment. In contrast, for rare-earth ions, the magnetic moments come from 4f electrons which are close to the inner core and have significant contributions from both spin and orbital angular momentum.[1] While the stronger interaction between the transition-metal ions determines the framework of the magnetic order in the RE-TM oxides[2–4], the weaker interaction between the rare-earth and transition-metal ions, on the other hand, generates interesting phenomena such as spin reorientations and moment compensation.[5–11] Despite the importance of the RE-TM interaction, a comprehensive understanding of its underpinnings and implications is still lacking for many material systems.

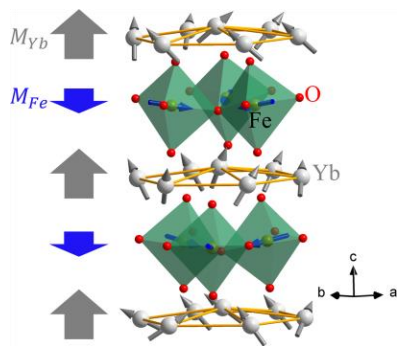

**Figure 1** (color online) The crystal structure of h-YbFeO$_3$ and schematic of the magnetic structure. The arrows on the atoms indicate the atomic magnetic moments. $M_{Fe}$ and $M_{Yb}$ are the magnetization of Fe and Yb along the $c$ axis respectively, which are anti-aligned at low temperature. The Fe moments form a 120-degree antiferromagnetic order in the basal plane, with only a very small component along the $c$ axis. The Yb moments are partially aligned by the Yb-Fe exchange field.

In this work, we study the magnetic interaction between the rare-earth and transition-metal ions by measuring the magnetization of the rare-earth and transition-metal ions separately using an element-specific method. In particular, we study hexagonal YbFeO$_3$, a member of hexagonal rare-earth ferrites (h-REFeO$_3$, RE=Ho-Lu, Y, and Sc). Hexagonal REFeO$_3$ have a layered crystal structure in which both RE and Fe atoms adopt a two-dimensional triangular lattice, as shown in Figure 1.[12] Below about 1000 K, h-REFeO$_3$ crystal structure undergoes a distortion corresponding to a rotation of the FeO$_5$ local structure and a buckling of the rare-earth layer, which induces an improper ferroelectricity.[13–17] The rotation of the FeO$_5$ also cants the moment on Fe, via the Dzyaloshinskii-Moriya interaction, generating weak ferromagnetism on top of a 120-degree antiferromagnetic order below about 120 K, as illustrated in Figure 1.[18–20] The spontaneous magnetization is along the $c$ axis. Recent work demonstrated that a super-lattice structure of hexagonal Lu-Fe-O materials are promising for realizing room temperature multiferroic materials with co-existing ferroelectricity and ferromagnetism,[21] a property that has potential application in energy-efficient information processing and storage[22].

In h-YbFeO$_3$, the Fe-Fe interaction is expected to dominate the framework of the magnetic ordering, as corroborated by the fact that the ordering temperature of h-YbFeO$_3$ is almost the same as that of h-LuFeO$_3$ (noting that Lu$^{3+}$ is non-magnetic)[16,17,23,24]. The Yb-Fe interaction is weaker but enough to partially align the moment on Yb and contribute to the total magnetization. Indeed, an enhancement of magnetization of h-YbFeO$_3$, compared with that in h-LuFeO$_3$, has been observed previously[23,24], to be up to about 3 $\mu_B$/f.u. at 3 K, in contrast to 0.018 $\mu_B$/f.u. in h-LuFeO$_3$.[13,16] The Yb-Fe interaction could, in principle, align or anti-align the moments of Fe and Yb. At the compensation temperature[3,5], the magnetization of Fe and Yb cancels, and an indication of this was observed previously at about 80 K[24]. On the other hand, direct observation of anti-alignment between the Fe and Yb magnetization is still lacking. In addition, the previously reported large magnetization (about 3 $\mu_B$/f.u.)[24] at low temperature is more consistent with a free Yb$^{3+}$, but unexpected when considering the effect of the crystal field generated by the local environment,[25–29] which could significantly change the effective magnetic moment and the magnetic anisotropy at low temperature.[28,30]

To elucidate the Yb-Fe interaction and the magnetic moment of Yb, we have studied the electronic structure of h-YbFeO$_3$ using X-ray absorption spectroscopy (XAS) and the X-ray photoemission spectroscopy (XPS) and measured the magnetization of Fe and Yb separately using X-ray magnetic circular dichroism (XMCD). We have found a large exchange field (17 kOe) on Yb, while the magnetic moment of Yb is significantly reduced from the value of a free ion. Mixed valence of Yb was investigated and found only at the surface of samples grown in reducing environment, suggesting minimal effect on the magnetism of h-YbFeO$_3$.

## II. Methods

Hexagonal YbFeO$_3$ (001) films (20-50 nm) were deposited on yttrium stabilized zirconia (YSZ) (111) substrates and on Fe$_3$O$_4$ (111)/Al$_2$O$_3$ (001) substrates using pulsed laser (248 nm) deposition in 5 mtorr oxygen and argon environment, at 750 °C with a laser fluence of about 1 J cm$^{-2}$ and a repetition rate of 2 Hz.[13,14,31] All the films studied with X-ray absorption spectroscopy and X-ray magnetic circular dichroism were grown in oxygen environment. The film growth was monitored using reflection high energy electron diffraction (RHEED). The crystal structures of the h-YbFeO$_3$ films were characterized by X-ray diffraction (XRD) using a Rigaku D/Max-B diffractometer, with the Co K-α radiation (1.7903 Å). The linear X-ray absorption spectroscopy on the Fe L edge and O K edge was studied using X-ray photoemission electron microscope (X-PEEM) at the SM beamline of the Canadian Light Source with linearly polarized X-ray. The circular X-ray absorption (fluorescence) spectroscopy of Yb M edge and Fe L edge measurements were performed at the bend magnet beamline 6.3.1 in the Advanced Light Source at Lawrence Berkeley National Laboratory and at the beamline 4IDC in the Advanced Photon Source at Argonne National Laboratory respectively. The angle-resolved X-ray photoemission spectra (ARXPS) were obtained using SPECS PHOIBOS 150 energy analyzer. A non-monochromatized Al Kα x-ray source, with photon energy 1486.6 eV, was used with various emission angles, as previously reported.[32]

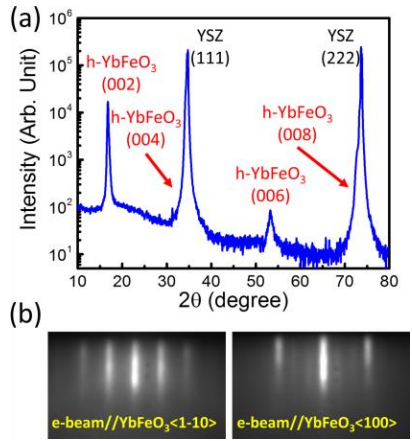

**Figure 2** (color online) (a) θ-2θ X-ray diffraction measurement of an h-YbFeO$_3$ film grown on yttrium stabilized zirconia (YSZ). (b) RHEED patterns of an h-YbFeO$_3$ film with electron beam along the <1-10> and <100> directions.

## III. Results and analysis

### A. Crystal structure and local environment of Fe

To verify the structure and phases of the epitaxial films, we carried out X-ray diffraction, electron diffraction, and X-ray spectroscopy measurements. Figure 2 (a) shows the X-ray diffraction (θ-2θ scan) of h-YbFeO$_3$/YSZ films. No additional peak other than those expected for h-YbFeO$_3$ and the substrate is visible in this large-range scan, indicating no impurity phases. As shown in Figure 2(b), RHEED images show diffraction streaks consistent with a flat surface and the structure of h-REFeO$_3$.[13,31]

The X-ray absorption spectra provided further confirmation of the local structure of Fe, from the Fe L edge

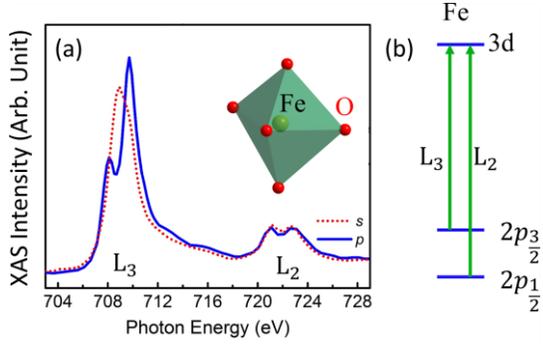

**Figure 3** (color online) (a) X-ray absorption spectra at the Fe L edge measured using linearly polarized X-ray. Inset: the FeO$_5$ local environment. (b) Schematic illustration of the L$_2$ and L$_3$ excitation.

spectra taken with linearly polarized X-ray. The local environment of Fe in h-YbFeO$_3$ is a trigonal bipyramid, with two apex O atoms (top and bottom) and three equator O atoms (in the Fe layer) as shown in Figure 1 as well as in Figure 3(a) inset. This structure makes the out-of-plane direction (along the *c* axis) and the in-plane direction (in the *a-b* plane) two distinct crystalline directions. Using linearly polarized X-ray, we measured the absorption spectra at the Fe L edge, as illustrated in Figure 3(b). As shown in Figure 3(a), the spectrum with *s*-polarized X-ray (*E* vector in the *a-b* plane) and that with *p*-polarized X-ray (*E* vector along the *c* axis) show obvious contrast, consistent with the large structural anisotropy. The spectra and linear dichroism in Figure 3(a) match those observed previously for h-LuFeO$_3$,[20,21,33,34] confirming that the local environment of the FeO$_5$ moiety in the two materials are almost identical.

## B. The Electronic structure of Yb

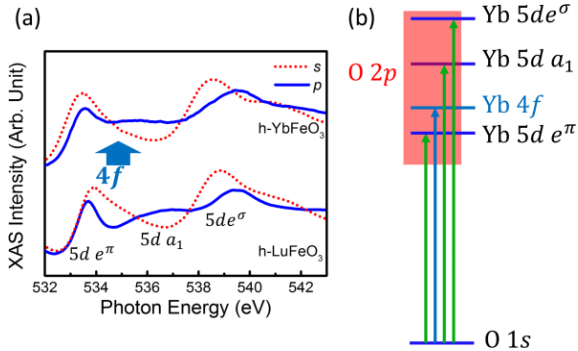

**Figure 4** (color online) (a) X-ray absorption spectra at the O K edge of h-LuFeO$_3$ and h-YbFeO$_3$ measured using linearly polarized X-ray. The arrow indicates the 4f state. (b) Schematic illustration of the O K edge excitation and the hybridization between the O and Yb states.

While the electronic structure of Fe in h-LuFeO$_3$ and h-YbFeO$_3$ are superficially similar, the electronic structure of Yb$^{3+}$ is expected to be different from that of Lu$^{3+}$ by one less 4f electron. To probe the unoccupied states of Yb, we measured the excitation of electrons from O 1s states to O 2p states (O K edge) using X-ray. Nominally, O 2p states are fully occupied; the O 1s to O 2p excitation is forbidden by the Pauli exclusion principle. If, on the other hand, the O 2p states are hybridized with the Yb states, the O 2p states will be slightly unoccupied and give rise to observable O 1s to O 2p excitation; one can infer the energy of the unoccupied Yb states using the excitation energies.[20] As shown in Figure 4(a), with linearly polarized X-rays, several features can be observed in the absorption spectra. Previously, we carried out symmetry analysis of the absorption spectra measured on h-LuFeO$_3$ and identified the origin of these features mainly as the 5d orbitals split in the crystal field: $e^\pi$, $a_1$ and $e^\sigma$ [see Figure 4(b)][20]. Compared with the X-ray absorption spectra of h-LuFeO$_3$, the spectra of h-YbFeO$_3$ show additional density of states, as indicated in Figure 4(a), which is expected to be the unoccupied 4f state that is hybridized with the O 2p states.

The 4f$^{13}$ configuration of Yb can also be probed by measuring the excitation directly to the unoccupied 4f states (in the absence s-f hybridization, none exist with Lu$^{3+}$). As shown in Figure 5(a), X-ray absorption spectra at the Yb M edge were measured at 18 K. Two peaks are observed in the absorption spectra at approximately 1513 and 1555 eV, which can be assigned to M$_5$ (initial state 3d$_{5/2}$) and M$_4$ (initial state 3d$_{3/2}$)

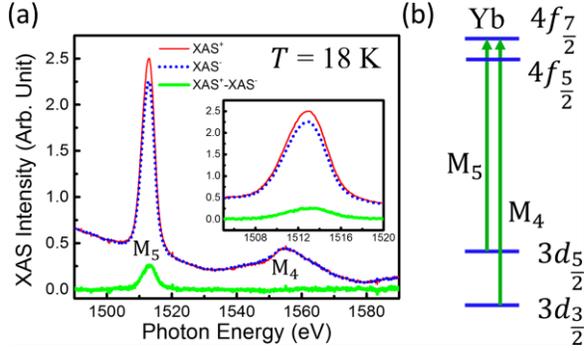

**Figure 5** (color online) (a) X-ray absorption spectra at the Yb M edge measured using X-ray polarized counterclockwise. XAS$^+$ (XAS$^-$) is the spectrum measured in magnetic field along the $+z$ ($-z$) direction. (b) Schematic illustration of the Yb M edge excitation. The crystalline $c$ axis of h-YbFeO$_3$ is along the $z$ direction.

excitations respectively according to the photon energy[35] [see Figure 5(b)]. The M$_5$ transition in Yb, which is allowed by the angular-momentum selection rule, can be described using the one-electron (hole) picture, without many-body interactions, due to the simple initial (full 3d$_{5/2}$, one hole in 4f$_{7/2}$) and final (one hole in 3d$_{5/2}$, full 4f$_{7/2}$) states, consistent with the observed sharp, structureless peak in Figure 5(a). The M$_4$ excitation (3d$_{3/2}$ to 4f$_{7/2}$), on the other hand, is not allowed by the angular-momentum selection rule. The non-zero intensity of the M$_4$ peak suggests that the crystal-field splitting and the Yb 4f-O 2p hybridization reduces the symmetry of the electronic states considerably, which is in line with the observed contribution to the O K edge excitation by the Yb 4f state shown in Figure 4(a).

## C. The Ferrimagnetism of h-YbFeO$_3$
### 1. Magnetization of Yb and Fe

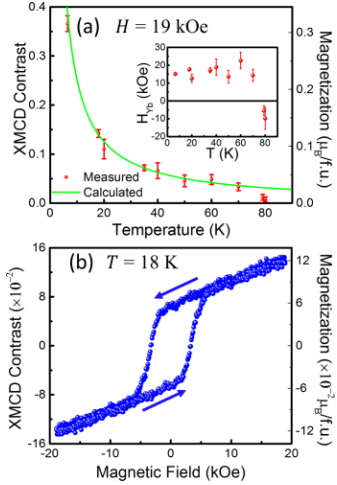

**Figure 6** (color online) XMCD contrast of Yb M$_5$ edge and the corresponding magnetization. (a) Temperature dependence measured in a 19 kOe magnetic field; the line is calculated using the parameters analyzed from (b). Inset: $H_{Yb}$ extracted from the mean-field theory (see text in Section IV.B). (b) Magnetic field dependence measured at 18 K. The magnetic field is along the $c$ axis.

To study the magnetization of Yb, we carried out X-ray magnetic circular dichroism measurements, by comparing the absorption spectra using circularly polarized X-ray in opposite magnetic fields. As shown in Figure 5(a), the X-ray absorption spectra measured in 19 kOe and -19 kOe magnetic fields along the $z$ direction show a clear contrast. We define the XMCD contrast as $\frac{I_+ - I_-}{(I_+ + I_-)/2}$, where $I_+$ and $I_-$ are the M$_5$ peak areas of the absorption spectra in positive and negative magnetic fields respectively.

The XMCD contrast measured at $H = 19$ kOe, for various temperatures between 6.5 and 80 K, is displayed in Figure 6(a). The value of the XMCD signal decreases rapidly at low temperature, inconsistent with typical ferromagnetic dependence, which typically follows the Bloch's law (decrease slowly at low temperature but much faster close to the magnetic ordering temperature)[36]. Figure 6(b) shows the field dependence of the XMCD contrast of Yb at 18 K. A clear hysteresis is observed with a coercive field of approximately 3.5 kOe. The magnetization converted from the XMCD contrast (See Appendix A) is also displayed in Figure 6.

Figure 7(a) shows the spectra of X-ray absorption of Fe L edge measured in circularly polarized X-ray in a 10 kOe magnetic field at 6.5 K. A clear difference is observed between the spectra measured using X-rays of different polarizations, which can be used to estimate the magnetization of Fe.[37] Figure 7(b) shows the magnetic-field dependence of the Fe magnetization calculated from the XMCD contrast using the sum rule[37–39]. A hysteretic behavior is observed, with a coercive field of approximately 4 kOe, consistent with the value found in previous bulk magnetometry measurements.[23,24] This coercive fields is also similar to that of Yb in Figure 6(b), indicative of the exchange field on Yb generated by Fe. The saturation magnetization of Fe is 0.05 +/- 0.01 µB/f.u., which corresponds to a small projection of the Fe moment along the $c$ axis. From Figure 6 and 7, we find that the magnetization of Fe is anti-parallel to the magnetic field and to that of the Yb magnetization at low temperature, as also illustrated in Figure 1. This provides a direct observation of ferrimagnetic order in h-YbFeO$_3$.

2. The Low temperature magnetic moment of Yb

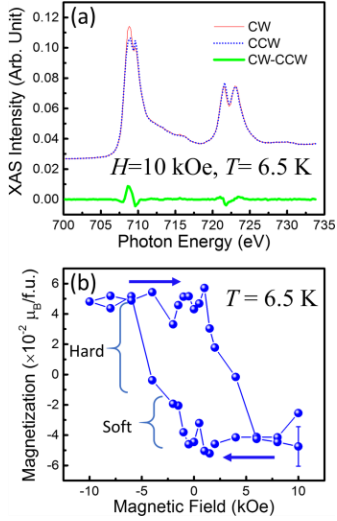

**Figure 7** (color online) (a) Absorption spectra of Fe L edge measured with circularly polarized X-ray in a 10 kOe field at 6.5 K. CW and CCW stand for clockwise and counterclockwise polarization of the X-ray respectively. (b) Magnetic-field dependence of the magnetization of Fe at 6.5 K, which contains a soft and a hard component (see discussion in Section IV.D). The magnetic field is along the $c$ axis.

As shown in Figure 6(b), the magnetization of Yb does not saturate in the measurement condition; instead, it shows a linear relation with magnetic field when the field is much larger than the coercive field, which is consistent with a susceptibility behavior and somewhat akin to paramagnetism for Yb. We can, nonetheless, further analyze the magnetic moment on Yb using the mean-field theory[36], which has been extensively discussed historically in orthoferrites and garnets[10,11,40–43].

In the mean-field theory, the exchange interactions are modeled using the molecular fields. Assuming that the saturation magnetization of Fe is $M_{Fe,S}$ (in µB/f.u.), the magnetization of Fe is given by:

$$M_{Fe} = M_{Fe,S}L(x_{Fe}), \quad (1)$$

where $L(x) = \coth(x) - \frac{1}{x}$ is the Langevin function, $x_{Fe} = \frac{(\gamma_{YbFe}M_{Yb} + \gamma_{Fe}M_{Fe} + \mu_0 H)M_{Fe,S}}{k_B T}$, $M_{Yb}$ is the magnetization of Yb, $\gamma_{YbFe}$ and $\gamma_{Fe}$ are the molecular field parameters for the Yb-Fe and Fe-Fe interactions respectively, $\mu_0$ is the vacuum permittivity, $k_B$ is the Boltzmann constant, $H$ is external magnetic field, and $T$ is temperature. The magnetization of Yb is given by:

$$M_{Yb} = \mu_{Yb}L(x_{Yb}), \quad (2)$$

where $x_{Yb} = \frac{(\gamma_{YbFe}M_{Fe} + \mu_0 H)\mu_{Yb}}{k_B T}$ and $\mu_{Yb}$ is the magnetic moment of Yb. No Yb-Yb exchange interaction is included since such exchange interactions are too weak to play a role in the temperature range investigated in this work.[3,5]

When the magnetic field is much larger than the coercive field and the temperature is much lower than the magnetic ordering temperature for the Fe ($\approx$ 120 K for h-YbFeO$_3$)[24], one may treat $|M_{Fe}| \approx M_{Fe,S}$ as a constant. As shown in Figure 6(b), at $T$ = 18 K, when $H$ is between 6 and 19 kOe, the XMCD contrast shows a linear dependence with magnetic field, suggesting that $\chi_{Yb}$ is small enough that the Langevin function takes a linear form with respect to the magnetic field $H$:

$$M_{Yb} = \frac{\mu_{Yb}^2(\gamma_{YbFe}M_{Fe} + \mu_0 H)}{3k_B T} \quad (3).$$

According to Eq. (3), the slope of the field dependence of $M_{Yb}$ (susceptibility) is $\chi_{Yb} = \frac{dM_{Yb}}{dH} = \frac{\mu_{Yb}^2 \mu_0}{3k_B T}$, which leads to $\mu_{Yb}$ = 1.6 +/- 0.1 $\mu_B$, a value much smaller than the magnetic moment of a free Yb (4.5 $\mu_B$/f.u.)[44].

3. Exchange field on Yb

According to Eq. (3), the remanent magnetization (magnetization in zero $H$) is expected to be

$$M_{Yb,R} = \frac{\mu_{Yb}^2 \gamma_{YbFe} M_{Fe}}{3k_B T} \quad (4).$$

Because $M_{Fe}$ and $M_{Yb}$ have different sign in zero $H$ [see Figure 6(b) and Figure 7(b)], one finds $\gamma_{YbFe} < 0$ from Eq. (4).

Using the value $M_{Yb,R}$ = 0.057 $\mu_B$/f.u. at 18 K from Figure 6(b), one can calculate the exchange field on Yb: $H_{Yb} = \frac{\gamma_{YbFe} M_{Fe}}{\mu_0}$ = 17 kOe. We also note that the exchange field on Yb generated by Fe in h-YbFeO$_3$ is about an order of magnitude larger than the value 1.6 kOe in orthorhombic YbFeO$_3$ and that in rare-earth orthoferrites in general[3]. This large difference may come from the dramatic differences between the bond lengths and bond angles in the hexagonal and orthorhombic YbFeO$_3$ structures.

D. The Possible mixed valence of Yb

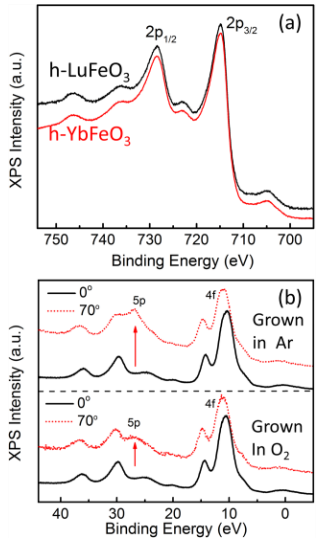

**Figure 8** (Color online) (a) The X-ray photoelectron spectra around Fe 2p edge of h-YbFeO$_3$ and h-LuFeO$_3$. (b) The X-ray photoelectron spectra around Yb 5p edge of h-YbFeO$_3$ film samples grown in Ar and O$_2$ environment measured at 0º and 70º take-off angle, corresponding to 2 nm and 0.7 nm probing depth respectively[47,48].

Mixed valence (Yb$^{3+}$ and Yb$^{2+}$) may play a role in the magnetism of h-YbFeO$_3$ as well as the determination of the magnetization on the Yb$^{3+}$. In principle, there is a tendency to form Yb$^{2+}$ due to the stability of the 4f$^{14}$ configuration. Although it will not affect the XMCD method discussed above since Yb$^{2+}$ does not contribute to the Yb M$_5$ X-ray absorption in the first place (the excitations to the fully occupied 4f states are forbidden in Yb$^{2+}$), it will be important for bulk magnetometry. We investigated the possibility of mixed valence in h-YbFeO$_3$ using ARXPS, by probing the core level electronic structure.

Figure 8(a) shows the Fe 2p X-ray photoemission spectra for both h-LuFeO$_3$ and h-YbFeO$_3$. The good match between the Fe 2p$_{3/2}$ peaks of h-LuFeO$_3$ and h-YbFeO$_3$ in Figure 8(a) indicates that Fe core level electronic structure are similar in these two ferrites. Previously, we have studied the X-ray photoemission spectra of Fe 2p using the Gupta and Sen (GS) multiplet fitting[45,46] of Fe 2p$_{3/2}$ in h-LuFeO$_3$ and concluded the Fe 2p and its satellite peaks are characteristic of a nominal Fe$^{3+}$ valance[32]. The same analysis applies here in h-YbFeO$_3$ as well. These features also do not vary with emission angle (data not shown). As a result, both the surface and the bulk part of the h-YbFeO$_3$ are in the nominal Fe$^{3+}$ valance state.

We also did not find indication of Yb$^{2+}$ in the film samples grown in oxygen environment (used for X-ray absorption spectroscopy and X-ray magnetic circular dichroism in Figure 2 to Figure 7). To investigate the possible appearance of Yb$^{2+}$, we studied ARXPS on the h-YbFeO$_3$ films prepared in argon environment A comparison with two samples grown in oxygen and argon environments is displayed in Figure 8(b). At the zero-degree take-off angle (perpendicular to surface), the XPS spectra of Yb are identical for both h-YbFeO$_3$ samples. At the 70° take-off angle, which probes mostly the surface[47,48], the XPS spectra of the sample grown in oxygen environment (lower panel) do not show clear difference from that at zero degree, also the surface appears to be slightly Yb rich. In contrast, for the sample grown in the argon environment, the XPS spectra at the 70° take-off angle exhibit additional intensity at the 5p peak, indicating a Yb$^{2+}$ valence[49] at the surface The correlation between the growth conditions indicates that the presence of oxygen vacancy promotes the reduction of Yb$^{3+}$. Although slightly YbO rich, the mixed surface termination (both iron oxide and YbO appear present at the surface) differs from the Fe–O termination seen for LuFeO$_3$.[32]

## IV. Discussion
### A. Origin of reduced moment of Yb

The low-temperature magnetic moment of Yb is found to be 1.6 µ$_B$, a value significantly smaller than 4.5 µ$_B$ for a free Yb [44]. In h-YbFeO$_3$, Yb is surrounded by 7 oxygen atoms, approximately corresponding to a C$_{3v}$ symmetry. Analysis using double groups indicates that the 4f$_{7/2}$ states are split by the crystal field into 4 levels: $3E_{\frac{1}{2}} + E_{\frac{3}{2}}$ (see Appendix B), where $E_{\frac{1}{2}}$ and $E_{\frac{3}{2}}$ are both two dimensional[44]. The energy scale of the crystal-field splitting is typically a few meV to a few tens meV,[26–28] which cannot be resolved in the XAS spectra. This crystal field splitting means that, at low temperature, only the low-lying level (ground state) is populated and contributes to the magnetization. The occupation of the low-lying level, in turn, leads to the reduced value of $\mu_{Yb}$, and is the reason for the temperature-dependent magnetic moments and magnetic anisotropy observed previously in rare-earth-containing oxides[28,30].

### B. Possible spin reorientation and magnetization compensation

One can calculate the temperature dependence of Yb magnetization using Eq. (2). As shown in Figure 6(a), the result (with $\mu_{Yb}$=1.6 µ$_B$, $H$=19 kOe, and $H_{Yb}$= 17 kOe) is compared with the measured values. The measured and the calculated magnetization match well below 70 K, suggesting that mean-field theory can describe the temperature dependence of $M_{Yb}$ too.. The fact that the mean-field theory can describe both the magnetic-field (Section III C 2) and temperature dependence of $M_{Yb}$, indicates its validity in analyzing the magnetic properties of h-YbFeO$_3$.

On the other hand, at about 80 K, the calculated value is much larger than the measured value, suggesting a reduction of $H_{Yb}$ at higher temperature. To reveal the temperature dependence of $H_{Yb}$, we calculated $H_{Yb}$ from the measured magnetization value using Eq. (3) (with $\mu_{Yb}$=1.6 µ$_B$, $H$=19 kOe); the result is displayed in Fig.6(a) inset. Clearly, a sign change of $H_{Yb}$ occurs at about 80 K, indicating a possible realignment between the magnetization $M_{Yb}$ and $M_{Fe}$, which is discussed below.

In principle, the alignment between $M_{Yb}$ and $M_{Fe}$ is determined by the minimization of total energy $E_{total} = -\frac{1}{2}\chi_{Yb}(H + \gamma_{YbFe}M_{Fe})^2 - M_{Fe}H$, or the maximization of the total magnetization $M_{total} = M_{Fe}(1 + \chi_{Yb}\gamma_{YbFe}) + \chi_{Yb}H$.[11] Here the external field $H$ is along the $c$ axis and $M_{Fe}$ may point either along or opposite to $H$, corresponding to the positive and negative signs respectively.

Because $\gamma_{YbFe} < 0$ and $\chi_{Yb} = \frac{\mu_{Yb}^2 \mu_0}{3 k_B T}$ (see section III C 3), the sign of $1 + \chi_{Yb}\gamma_{YbFe}$ is expected to change with temperature, possibly causing the reversal of the direction of the magnetization $M_{Fe}$:

(1) At low temperature, $1 + \chi_{Yb}\gamma_{YbFe} < 0$. In this case, $M_{Fe} < 0$ ($M_{Fe}$ anti-parallel to $H$) is more favorable for maximizing $M_{total}$. this mean the exchange field $H_{Yb} = \frac{\gamma_{YbFe}M_{Fe}}{\mu_0} > 0$ (parallel to the external field).

(2) When temperature is increased and $1 + \chi_{Yb}\gamma_{YbFe} > 0$ is satisfied, $M_{Fe} > 0$ ($M_{Fe}$ parallel to $H$) is more favorable. In this case, the exchange field $H_{Yb} = \frac{\gamma_{YbFe}M_{Fe}}{\mu_0} < 0$ (antiparallel to the external field); this could be the reason that at about 80 K $H_{Yb}$ becomes negative [Figure 6(a) inset]. .

(3) At the compensation temperature, $1 + \chi_{Yb}\gamma_{YbFe} = 0$. Therefore, $M_{total} = \chi_{Yb}H$, as if $M_{Fe}$ is "screened" by the part of Yb moment induced by the exchange field $H_{Yb}$. The magnetization compensation can be understood as the cancellation of $M_{Fe}$ and $M_{Yb}$ at zero field. According to Fig. 6(a) inset, the compensation temperature appears to be between 70 and 80 K, in fair agreement with the previous estimation[24].

Nonetheless, the magnetization of the Yb is largely a spectator to that of the Fe. The coercivity is the same as that observed for iron, with the essential observation [Figure 6(b)] that the magnetization does not easily saturate indicating that much of the magnetization depends on the magnetic susceptibility and possible alignment of the moments with external magnetic field $H$ and with that of Fe (Figure 7).

### C. Exchange field on Fe

The exchange field may also have an effect on the Fe, which can be understood by combining Eq. (1) and Eq. (2) to reach $x_{Fe} = \frac{\left[\gamma_{YbFe}\mu_{Yb}L\left(\frac{\gamma_{YbFe}M_{Fe}\mu_{Yb}}{k_B T}\right) + \gamma_{Fe}M_{Fe}\right]M_{Fe,S}}{k_B T}$, assuming $H = 0$. Since Fe moments in h-YbFeO$_3$ form weak ferromagnetic order, $\gamma_{Fe}$ must be positive. Because of the properties of the Langevin function $L(x)$, $\gamma_{YbFe}\mu_{Yb}L\left(\frac{\gamma_{YbFe}M_{Fe}\mu_{Yb}}{k_B T}\right)$ is always positive regardless of the sign of $\gamma_{YbFe}$. Therefore, the Yb always enhances the molecular field on the Fe. That said, because in general $\gamma_{Fe} \gg |\gamma_{YbFe}|$, the effect may not be significant.

### D. Comparison between magnetic properties of h-YbFeO$_3$ and h-LuFeO$_3$

Hexagonal LuFeO$_3$ (h-LuFeO$_3$) is the most studied hexagonal rare-earth ferrites. Because Lu$^{3+}$ is non-magnetic, the magnetic properties of h-LuFeO$_3$ is less complex. By comparing h-LuFeO$_3$ and h-YbFeO$_3$, one may gain insight on the effect of the rare earth on the magnetism.

One dramatic difference between h-YbFeO$_3$ and h-LuFeO$_3$ is in the coercive field of magnetization. For h-YbFeO$_3$ at 18 K, the coercive field is about 4 kOe, which is much smaller than the value 25 kOe for h-LuFeO$_3$.[16] For both h-LuFeO$_3$ and h-YbFeO$_3$, the magnetization-field loops for Fe have a squared shape, suggesting that the magnetic coercive field is determined by the competition between the magnetic anisotropy energy and the Zeeman energy. Compared with h-LuFeO$_3$, h-YbFeO$_3$ has enhanced magnetization due to the contribution of Yb. Therefore, a much smaller magnetic field is needed in h-YbFeO$_3$ to overcome the magnetic anisotropy, corresponding to a much smaller coercive field.

Another difference between h-YbFeO$_3$ and h-LuFeO$_3$ is in the saturation magnetization of Fe. According to Figure 7, in h-YbFeO$_3$, $M_{Fe,S} = 0.05$ +/- 0.01 $\mu_B$/f.u., larger than that in h-LuFeO$_3$ ($\approx 0.03$ $\mu_B$/f.u.)[16]. We note that previously it was observed in h-LuFeO$_3$ that the magnetization contains a soft component and a hard component, in which only the hard component (0.018 $\mu_B$/f.u.) is believed to be intrinsic to the weak ferromagnetic ordering because it disappears above the magnetic ordering temperature.[16] In Figure 7, there is also one soft (coercive field $\approx 1$ kOe) and one hard component (coercive field $\approx 4$ kOe). If we only treat the hard component to be intrinsic to the canting of the Fe moment, the weak ferromagnetic moment of Fe in h-YbFeO$_3$ is = 0.03 +/- 0.01 $\mu_B$/Fe according to Fig. 7(b), still larger compared with the value 0.018 $\mu_B$/f.u. in h-LuFeO$_3$.[16] Due to the size difference of Lu$^{3+}$ and Yb$^{3+}$,[14] the lattice constant of the basal plane of h-LuFeO$_3$ is smaller than that of h-YbFeO$_3$: $a = 5.963$ Å for h-LuFeO$_3$ and $a = 6.021$ Å for h-YbFeO$_3$.[31] Our recent work suggests that a compressive biaxial strain may reduce the canting of the Fe moments in h-REFeO$_3$,[50] which is in line with the correlation between the lattice constant and weak ferromagnetic moment on Fe observed here.

# V. Conclusion

We have studied the electronic structure and magnetic ordering of h-YbFeO$_3$ (001) thin films on YSZ (111) and on Fe$_3$O$_4$(111)/Al$_2$O$_3$(001) substrates. The magnetism of Yb in h-YbFeO$_3$ was studied using the element-specific method X-ray magnetic circular dichroism based on X-ray absorption spectroscopy. From the temperature and magnetic-field dependence of the Yb magnetization, we found that the low temperature Yb magnetic moment is significantly reduced compared with the value of free Yb$^{3+}$ ions, indicating the effect of crystal field. The exchange field on Yb, generated by the Fe moments, tends to anti-align the magnetization of Fe and Yb at low temperature. We also investigated possible valence mixing of Yb and only found indication of Yb$^{2+}$ at the surface of samples grown in an Ar environment, suggesting an insignificant effect to the magnetism of h-YbFeO$_3$. We expect that future work, such as optical spectroscopy on probing Yb crystal field levels and theoretical calculations on Yb-Fe interaction strength, may provide more insight on the ferrimagnetism of h-YbFeO$_3$.

# Acknowledgements

This project was primarily supported by the National Science Foundation through the Nebraska Materials Research Science and Engineering Center (Grant No. DMR-1420645). Additional support was provided by the Semiconductor Research Corporation through the Center for Nanoferroic Devices and the SRC-NRI Center under Task ID 2398.001. Work of Bryn Mawr College is supported by National Science Foundation Career Award (Grant No. NSF DMR-1053854). This research used resources of the Advanced Photon Source, a U.S. Department of Energy (DOE) Office of Science User Facility operated for the DOE Office of Science by Argonne National Laboratory under Contract No. DE-AC02-06CH11357. Use of the Advanced Light Source was supported by the U.S. Department of Energy, Office of Science, Office of Basic Energy Sciences under contract no. DE-AC02-05CH11231. The Canadian Light Source is funded by the Canada Foundation for Innovation, the Natural Sciences and Engineering Research Council of Canada, the National Research Council Canada, the Canadian Institutes of Health Research, the Government of Saskatchewan, Western Economic Diversification Canada, and the University of Saskatchewan. The research was performed in part in the Nebraska Nanoscale Facility: National Nanotechnology Coordinated Infrastructure and the Nebraska Center for Materials and Nanoscience, which are supported by the National Science Foundation under Award ECCS: 1542182, and the Nebraska Research Initiative.

# Appendix

## A. Converting XMCD contrast to magnetization of Yb

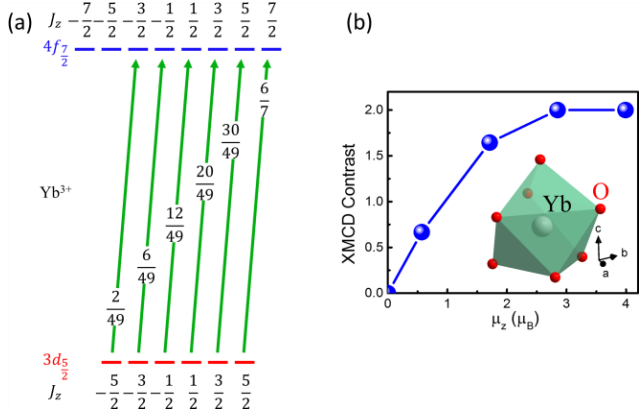

**Figure 9** (color online) (a) Transition probability between individual $3d_{\frac{5}{2}}$ and $4f_{\frac{7}{2}}$ states excited by a clockwise polarized X-ray. (b) XMCD contrast as a function of $\mu_z$ (see text) calculated assuming a free $Yb^{3+}$ ion. The inset shows the local environment of Yb with a $C_{3v}$ symmetry.

The calculation of the magnetization of Yb from the XMCD contrast at the M edge is different from that of 3d metals (e.g. Fe, Co, Ni) at the L edge, due to the strong spin-orbit coupling in both initial and final states. We hereby present a method based on the XMCD contrast of excitations from the $3d_{5/2}$ to the individual $4f_{7/2}$ eigenstates $J_z=-7/2$ to $7/2$, where $J_z$ is the projection of total angular moment $J$ on the $z$ axis; all possible $4f_{7/2}$ states are superposition of these states.

Excited by an X-ray polarized clockwise, the transition from one $3d_{5/2}$ state to one $4f_{7/2}$ state needs to satisfy $\Delta J_z = 1$. One can calculate the transition probabilities $P$ between individual states; the non-zero results are displayed in Figure 9(a). The projection of the magnetic moment of a $J_z$ state on the $z$ direction is $\mu_z = g\mu_B J_z$, where $g = 1.14$ is the Lande $g$-factor and $\mu_B$ is the Bohr magneton. Therefore, one can calculate the XMCD contrast, defined as $\frac{2(P_{J_z}-P_{-J_z})}{P_{J_z}+P_{-J_z}}$, with respect to $\mu_z$, where $P_{J_z}$ ($P_{-J_z}$) is the transition probability for the final state represented by $J_z$ (-$J_z$); the result is shown in Figure 9(b). Although for large $\mu_z$, the XMCD contrast does not distinguish the $|J_z|=7/2$ and $|J_z|=5/2$ states, for small $\mu_z$, the relation between XMCD contrast and $\mu_z$ is approximately linear. The measured XMCD contrast in this work falls in the small $\mu_z$ region (all values are less than 0.4). Therefore, we can use the relation in Fig. 9(b) to convert XMCD contrast to magnetization as a fair approximation.

## B. Group theory analysis of the crystal field splitting of Yb states

In h-YbFeO$_3$, the local environment of Yb has a symmetry that can be described using point group $C_{3v}$ [see Figure 9(b) inset]. The degenerate electronic states in general are split according to the symmetry of the local environment. Because of the strong spin-orbit coupling, the angular momentum of the 4f states takes half-integer $J = \frac{5}{2}$ or $J = \frac{7}{2}$, the analysis of which requires the double group. TABLE 1 shows the character table for $C_{3v}$ double group, including irreducible representations $A_1, A_2, E, E_{\frac{1}{2}}, E_{\frac{3}{2}}$ ($E_{\frac{3}{2}+}$ and $E_{\frac{3}{2}-}$). The character of the representation with angular momentum $J = \frac{5}{2}$ and $J = \frac{7}{2}$ are also listed. Using these characters, one can reduce the $J = \frac{5}{2}$ and $J = \frac{7}{2}$ representations. The results are: $J = \frac{5}{2} \rightarrow 2E_{\frac{1}{2}} + E_{\frac{3}{2}}$ and $J = \frac{7}{2} \rightarrow 3E_{\frac{1}{2}} + E_{\frac{3}{2}}$.

TABLE 1 Character table of the double group $C_{3v}$

| $C_{3v}$ | E | $2C_3$ | $3\sigma_v$ | RE | $2RC_3$ | $3R\sigma_v$ |
|---|---|---|---|---|---|---|
| $A_1$ | 1 | 1 | 1 | 1 | 1 | 1 |
| $A_2$ | 1 | 1 | -1 | 1 | 1 | -1 |
| $E$ | 2 | -1 | 0 | 2 | -1 | 0 |
| $E_{\frac{1}{2}}$ | 2 | 1 | 0 | -2 | -1 | 0 |
| $E_{\frac{3}{2}^+}$ | 1 | -1 | i | -1 | 1 | -i |
| $E_{\frac{3}{2}^-}$ | 1 | -1 | -i | -1 | 1 | i |
| $J = \frac{7}{2}$ | 8 | 1 | 0 | -8 | -1 | 0 |
| $J = \frac{5}{2}$ | 6 | 0 | 0 | -6 | 0 | 0 |